\documentclass[prb,superscriptaddress,letter,reprint]{revtex4-1}
\usepackage{graphicx}
\usepackage[unicode=true,colorlinks=true,urlcolor=blue,citecolor=blue]{hyperref}

\begin{document}

\title{Sign-reversal electron magnetization in Mn-doped semiconductor
structures}

\author{I. A. Kokurin}
\email[E-mail:]{kokurinia@math.mrsu.ru} \affiliation{Ioffe
Institute, 194021 St. Petersburg, Russia} \affiliation{Institute of
Physics and Chemistry, Mordovia State University, 430005 Saransk,
Russia} \affiliation{St. Petersburg Electrotechnical University
``LETI,'' 197376 St. Petersburg, Russia}
\author{A. Yu. Silov}
\affiliation{Department of Applied Physics and Institute for
Photonic Integration, Eindhoven University of Technology, P.O. Box
513, NL-5600MB Eindhoven, The Netherlands}

\author{N. S. Averkiev}
\affiliation{Ioffe Institute, 194021 St. Petersburg, Russia}

\begin{abstract}
The diversity of various manganese types and its complexes in the
Mn-doped ${\rm A^{III}B^V}$ semiconductor structures leads to a
number of intriguing phenomena. Here we show that the interplay
between the ordinary substitutional Mn acceptors and interstitial Mn
donors as well as donor-acceptor dimers could result in a reversal
of electron magnetization. In our all-optical scheme the
impurity-to-band excitation via the Mn dimers results in direct
orientation of the ionized Mn-donor $d$ shell. A photoexcited
electron is then captured by the interstitial Mn and the electron
spin becomes parallel to the optically oriented $d$ shell. That
produces, in the low excitation regime, the spin-reversal electron
magnetization. As the excitation intensity increases the capture by
donors is saturated and the polarization of delocalized electrons
restores the normal average spin in accordance with the selection
rules. A possibility of the experimental observation of the electron
spin reversal by means of polarized photoluminescence is discussed.
\end{abstract}

\date{\today}

\maketitle

\textit{Introduction.} The control of a spin state and the related
magnetization of the charge carriers by nonmagnetic methods is a key
concept of the semiconductor spintronics. A reversal of the
magnetization by electrical or optical means may help construct the
low-power spintronic devices eliminating the conventional magnetic
switching method. The manipulation of magnetization by the electric
field or current is well known both in nonmagnetic \cite{Silov2004}
and in magnetic semiconductors (e.g., GaMnAs)
\cite{Matsukura2015,Lee2017} and in hybrid
semiconductor/ferromagnetic structures \cite{Tivakornsasithorn2018}.

The absorption of the circularly polarized light leads to the spin
polarization of the nonequilibrium carriers in the semiconductor
structures due to spin-orbit interaction. Thus, the optical
orientation~\cite{Meier1984,Dyakonov2008} is simply the conservation
of angular momentum in a system of the electrons and the photons. It
is well known that the optical selection rules strictly couple a
photon polarization with the electron spin state during the
photoexcitation. However, in a steady state it is necessary to take
into account not only the excitation processes (selection rules) but
also the relaxation processes as well.

Usually the spin state of nonequilibrium charge carriers is
determined experimentally by means of the polarized
photoluminescence (PL). The so-called ``negative'' PL polarization
(the PL polarization helicity is opposite to that of an excitation)
does not necessarily indicate the reorientation of the photoexcited
spin. Moreover, this frequently corresponds to the recombination of
the resident carrier whose spin is aligned due to the exchange
interaction with an exciton (see, for instance,
Ref.~\onlinecite{Ignatiev2009}). Unlike such processes here we
suggest a mechanism of the anomalous alignment of the nonequilibrium
carrier spin.

A possibility to address individual impurities \cite{Koenraad2011}
by optical methods holds considerable promise
\cite{Leger2006,Kudelski2007}. The direct manipulation of the
impurity spin via the impurity-to-band excitation (the
photoneutralization transition) has an advantage in that respect
compared to the band-to-band one. The optical transitions involving
impurities are well known from the early 1960s
\cite{Eagles1960,Dumke1963}. The possible use of these transitions
for the optical orientation was recently
discussed~\cite{Kokurin2013} and experimentally
demonstrated~\cite{Petrov2016}.

In this communication a possibility to change the magnitude and the
sign of the electron magnetization in the Mn-doped GaAs structures
by optical means alone is foretold. A model describing a possibility
to govern the electron spin magnetization by utilizing the
impurity-to-band excitation scheme is proposed. The magnetization is
described by a system of rate equations involving the conduction
band electrons and various Mn complexes. We also consider an
experimental implementation of our findings.

\textit{Rate equations and electron sign-reversal magnetization.} --
Our model takes into account the following Mn complexes. The
conventional Mn acceptor substituting Ga-cation in GaAs lattice (see
a recent review \cite{Averkiev2018}, and references therein) is
referred to as ${\rm Mn_{Ga}}$. Double donors ${\rm Mn_I}$ arising
in the interstitial positions are taken into consideration as
well~\cite{Maca2002,Yu2002}. It is essential that the closely spaced
ionized donors and acceptors can form pairs or (${\rm Mn_I}$-${\rm
Mn_{Ga}}$)-dimers \cite{Dietl2014}.

Let us consider the optical orientation through the
photoneutralization transition (excitation from an ionized acceptor
state to the conduction band) in Mn-doped GaAs structures. Such a
transition, ${\rm Mn^-_{Ga}}+\hbar\omega\rightarrow {\rm
Mn^0_{Ga}}+e^-$, with the excitation energy $\hbar\omega$ less than
a band gap in the case of deep acceptors (${\rm Mn^0_{Ga}}$
ionization energy in bulk GaAs is about 110 meV) can be realized in
compensated structures only \footnote{Using the structures with the
quantum wells has an additional advantage of the complementary
compensation from the donors in the barriers of the modulation-doped
structures.}. The presence of various Mn types and complexes and
accounting of exchange interaction specifics therein provides the
possibility for the electron sign-reversal magnetization (SRM) in
the system of photoexcited electrons.

\begin{figure}
\includegraphics[width=57mm]{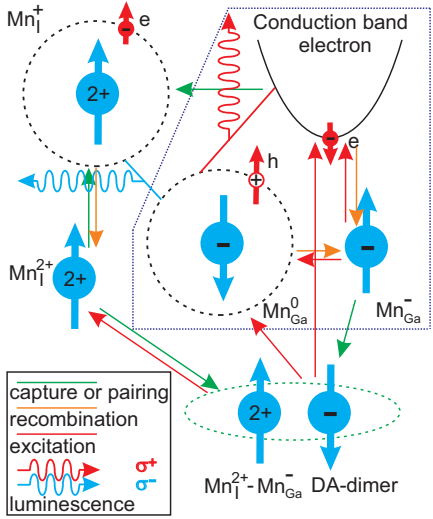}
\caption{\label{fig01} The kinetic scheme for the SRM description.
Transitions between the following charge states and complexes are
included in kinetics: ${\rm Mn^-_{Ga}}$ ionized acceptor, ${\rm
Mn^0_{Ga}}$ neutral acceptor, ${\rm Mn^{2+}_I}$ and ${\rm Mn^+_I}$
ionized donors, (${\rm Mn^{2+}_I}$-${\rm Mn^-_{Ga}}$) dimer as well
as delocalized conduction band electrons. The wide arrows indicate
the predominant spin orientation in each state. Thin arrows
correspond to different processes and transitions between charge
states and complexes. The photoexcitation, spin relaxation,
radiative recombination, carrier capture by ionized donors, and
Mn-dimer formation are taken into account.}
\end{figure}

The kinetic scheme for SRM is shown in Fig.~\ref{fig01}. The
above-mentioned configurations of the manganese and their different
charged states are taken here into account. In equilibrium there are
both acceptors and donors in the ionized states ${\rm Mn_{Ga}^-}$
and ${\rm Mn_I^{2+}}$, respectively. Their relative concentration
depends on the compensation degree. Both of these charge states will
be classified by the total $d$ shell spin of $S=5/2$. The
interstitial Mn donor can also capture an extra electron forming a
very shallow donor state ${\rm Mn^+_I}$~\cite{Dietl2014}. The
exchange interaction of the electron and the $d$ shell is
ferromagnetic in this case (the ground state with the total angular
momentum of 3). The exchange interaction of the $d$ shell and the
acceptor bound hole (the angular momentum $J=3/2$
\cite{Gelmont1972,Baldereschi1973}) is antifferomagnetic, which
leads to the ground state of ${\rm Mn^0_{Ga}}$ with the total
angular momentum $F=1$ \cite{Averkiev2018}. The exchange interaction
between two $d$ shells in (${\rm Mn_I^{2+}}$-${\rm Mn_{Ga}^-}$)
dimer is antiferromagnetic \cite{Blinowski2003,Masek2004}, which
corresponds to the ground state with the zero magnetic moment.

The pseudospin-1/2 model is utilized to describe both ${\rm
Mn_{Ga}}$ and ${\rm Mn_I}$ states, which is a sufficient scheme to
explain the SRM phenomenon. The following concentrations notation is
used in our rate-equation model. Here the thick arrows $\Uparrow$
($\Downarrow$) correspond to the $+5/2$ ($-5/2$) projection of $d$
shell spin. By analogy $\uparrow$ ($\downarrow$) indices point out
onto electrons or holes with the spin projection $+1/2$ ($-1/2$) and
$+3/2$ ($-3/2$), respectively. For the conduction band electrons
$n_\uparrow$ ($n_\downarrow$) are assigned to the spin-up
(spin-down) state. In the ground state of ${\rm Mn^0_{Ga}}$ we take
into account $m^0_{\Uparrow\downarrow}$ ($m^0_{\Downarrow\uparrow}$)
for states with $+1=+5/2-3/2$ ($-1=-5/2+3/2$) angular momentum
projection. The singly ionized ${\rm Mn^+_I}$ donor concentrations
with angular momentum projections $+3=5/2+1/2$ ($-3=-5/2-1/2$) are
labeled by $m^+_{\Uparrow\uparrow}$ ($m^+_{\Downarrow\downarrow}$).
The concentrations $m^-_\Uparrow$ ($m^-_\Downarrow$) correspond to
the states of ${\rm Mn^-_{Ga}}$ with the $d$-shell spin $+5/2$
($-5/2$). Similarly $m^{2+}_\Uparrow$ ($m^{2+}_\Downarrow$)
describes the ${\rm Mn^{2+}_I}$ donor state with the $d$-shell spin
$+5/2$ ($-5/2$). Finally, $m^d$ stands for the (${\rm
Mn_I^{2+}}$-${\rm Mn_{Ga}^-}$)-dimer ground state concentration.
This leads to the system of $5\times 2+1=11$ rate equations
\cite{Suppl}. Each ${\rm Mn_I^{2+}}$ donor is supposed in
equilibrium to be a part of the dimer.

Two separate channels of excitation are considered in our scheme
(see Fig.~\ref{fig01}). The first one is the photoneutralization of
the ionized acceptor ${\rm Mn_{Ga}^-}$, and the second one
corresponds to the photoneutralization of the ionized acceptor
inside the (${\rm Mn_I^{2+}}$-${\rm Mn_{Ga}^-}$) dimer. The
$\sigma^+$ excitation acts on the $\Downarrow$ state of ionized
acceptors. In the former case the conduction band electron in the
$\downarrow$ state and the neutral acceptor ${\rm Mn^0_{Ga}}$ with
$d$-shell--hole spin configuration $\Downarrow\uparrow$ ($F_z=-1$)
arise. Furthermore, this leads to predominant
$\Uparrow$-polarization of the remaining ionized acceptors. In the
latter case an ionized manganese ${\rm Mn^{2+}_I}$ additionally
appears with $\Uparrow$-polarized $d$ shell. The above spin
configurations are determined by optical selection rules and the
antiferromagnetic alignment between an ionized donor and acceptor
inside the Mn dimer. The predominant $\Uparrow$ orientation of ${\rm
Mn^{2+}_I}$ leads to $\Uparrow\uparrow$ spin configuration of ${\rm
Mn^+_I}$ after the electron capture by a magnetic donor due to the
ferromagnetic alignment.

It is well known that the generation term has the form $\alpha G$
\cite{Ryvkin1964}, where $\alpha$ is the absorption coefficient and
$G$ is the photon flux density. However, for the impurity-assisted
transition the absorption coefficient is proportional to the
impurity concentration \cite{Eagles1960}. We denote the absorption
coefficient $\alpha^-$ for the ${\rm Mn^-_{Ga}}$ concentration
$M^-$. By analogy the absorption coefficient $\alpha^d$ corresponds
to the Mn-dimer concentration $M^d$. The concentration-independent
quantity is $\alpha^-/M^-$ ($\alpha^d/M^d$), that is, the absorption
cross section. It should be noted that the concentration $M^-$
depends not only on a compensation from ${\rm Mn_I}$, but can
additionally increase due to the residual donors with concentration
$M^{\rm back}$ excluded from kinetics, $M^-=M^d+M^{\rm back}$.

Within our pseudospin-1/2 model the electrons in the spin-up and the
spin-down state recombine with the hole of the ${\rm Mn^0_{Ga}}$
acceptor with the emission of $\sigma^-$- and $\sigma^+$-polarized
light, respectively. The intensity of the bimolecular recombination
is described by $\gamma^{\rm eA}$ and $\gamma^{\rm DA}$ coefficients
for the electron-acceptor (eA) and donor-acceptor (DA)
recombination, respectively. The capture rate of the photoexcited
electrons by the magnetic donors is described by the $\beta^{\rm
sc}$ ($\beta^{\rm sf}$) coefficient for the process with the spin
conservation (spin flip). We consider the dimer formation from
single ${\rm Mn^{2+}_I}$ and ${\rm Mn^-_{Ga}}$, which depends on the
$d$-shell spin state of the donor and acceptor. The dimer formation
rate is determined by the coefficient $\delta^{\rm sc}$ for the
process with the conservation of $d$-shell spins. By analogy the
parameter $\delta^{\rm sf}$ corresponds to the spin-flip process.
Here ${\rm Mn^{2+}_I}$ and ${\rm Mn^-_{Ga}}$ $d$-shell spins are
supposed to be uncorrelated after DA recombination (${\rm
Mn_I^+}+{\rm Mn_{Ga}^0}\rightarrow{\rm Mn_I^{2+}}+{\rm
Mn_{Ga}^-}+\hbar\omega$), i.e., the direct DA-dimer formation during
the DA recombination is neglected. For this there are two reasons:
(i) the electron of ${\rm Mn^+_I}$ donor can recombine with the hole
of ${\rm Mn^0_{Ga}}$ center that was not a part of the DA dimer;
(ii) the spin configuration of $d$ shells of the recombining Mn
donor and acceptor is ferromagnetic, which prevents the dimer
formation in the ground state. The spin relaxation is taken into
account in each charge state with a nonzero angular momentum.

In general, the steady state of the system is described by the
system of 11 nonlinear algebraic equations~\cite{Suppl}. This
problem can be solved only numerically. However, the SRM effect can
be qualitatively demonstrated by solving the rate equations in
limiting cases of low and high pump power analytically.

At the low pump power, the excitation accompanied by the Mn-dimer
decay is dominant, which is the consequence of a higher absorption
coefficient \footnote{${\rm Mn_{Ga}}$ in the immediate vicinity of
the ionized ${\rm Mn_I}$ provides a better spatial overlap of the
bound hole and the electrons in the conduction band. In the context
of the reciprocity of the absorption and the emission, it leads to a
higher absorption coefficient in the DA-dimer state compared to the
single ionized acceptor.}. Due to the low concentration of
recombining carriers in this regime, the radiative lifetime $\tau_l$
is the longest timescale. In this case the capture by ionized donors
is the main mechanism for the decrease of the electron
concentrations $n_\uparrow$ ($n_\downarrow$). The dominance of the
capture mechanism with the electron spin conservation or spin flip
depends on the relation between the electron spin relaxation time
$\tau_s^e$ and the capture time $\tau_c$ and is governed by the pump
power, since the capturing rate is proportional to the ionized donor
concentration. In the former case the fast spin relaxation is
realized for conduction band electrons, $\tau_s^e\ll\tau_c$. This
corresponds to the capture of relaxed $\uparrow$-polarized electrons
by $\Uparrow$-polarized ${\rm Mn^{2+}_I}$ donors. In the latter case
we have the opposite situation, $\tau_s^e\gg\tau_c$. Then
predominantly $\downarrow$-oriented electrons have to reverse their
spin during the capture by polarized ${\rm Mn^{2+}_I}$ (the donor
$d$-shell spin-relaxation time $\tau_s^{2+}$ is supposed to be long
as well). In both cases the spin of the donor-bound electron is
oriented oppositely to that of delocalized electrons at the
excitation moment.

Thus, in the low pump power regime, the ${\rm Mn^+_I}$ state with
the angular momentum projection of $+3$ predominantly arises (see
the left side of the diagram in Fig.~\ref{fig01}). Here all excited
electrons are supposed to be captured by the donors into the
$\Uparrow\uparrow$ state of ${\rm Mn^+_I}$. This corresponds to the
direct generation term in the rate equation describing the
population of the ${\rm Mn_I^+}$ $\Uparrow\uparrow$ state. Here the
capture time $\tau_c$ (or any terms containing capture parameters
$\beta^{\rm sc}$, $\beta^{\rm sf}$) is not included in the
simplified system of rate equations describing ${\rm Mn^+_I}$
states. Finally, the above scheme can be described by two simplified
rate equations for ${\rm Mn^+_I}$-donor concentration:

\begin{equation}
\label{m_up}
\frac{dm^+_{\Uparrow\uparrow}}{dt}=\frac{\alpha^d}{M^d}m^dG-\gamma^{\rm
DA}m^+_{\Uparrow\uparrow}m^0_{\Uparrow\downarrow}-\frac{m^+_{\Uparrow\uparrow}-m^+_{\Downarrow\downarrow}}{2\tau_s^+},
\end{equation}
\begin{equation}
\label{m_down}
\frac{dm^+_{\Downarrow\downarrow}}{dt}=-\gamma^{\rm
DA}m^+_{\Downarrow\downarrow}m^0_{\Downarrow\uparrow}-\frac{m^+_{\Downarrow\downarrow}-m^+_{\Uparrow\uparrow}}{2\tau_s^+}.
\end{equation}

Here $\tau_s^+$ is the ${\rm Mn_I^+}$ spin-relaxation time. The
steady-state solution of these equations in the limit of the fast
${\rm Mn_{Ga}^0}$ spin relaxation and the total compensation
($m^0_{\Uparrow\downarrow}=m^0_{\Downarrow\uparrow}=m^0/2=m^+/2$)
gives for the donor concentration
$m^+=m^+_{\Uparrow\uparrow}+m^+_{\Downarrow\downarrow}$ and for a
difference in population of the opposite spin states $\Delta
m^+=m^+_{\Uparrow\uparrow}-m^+_{\Downarrow\downarrow}$ the following
result:
\begin{equation}
m^+=\sqrt{\frac{2\alpha^dG}{\gamma^{\rm DA}}},\qquad \Delta
m^+=\alpha^d\tau_s^+G,
\end{equation}
corresponding to the anomalous magnetization compared to the
spin-down polarization under the optical orientation in GaAs. The
fast ${\rm Mn^0_{Ga}}$ spin relaxation is not a necessary condition
to observe the anomalous electron magnetization. Here this
assumption is used to reduce the number of variables in equations.
On the contrary, the growing ${\rm Mn^0_{Ga}}$ spin-relaxation time
makes the SRM phenomenon more pronounced (anomalous magnetization
grows), which is due to the deceleration of the DA recombination.

In the opposite case of a high pump power the excitation from ${\rm
Mn_{Ga}^-}$ is dominant due to a saturation of the channel with
donors~\footnote{The donor concentration $M^d$ does not exceed 10\%
of the total Mn concentration $M$ \cite{Dietl2014}.}. Thus, mainly
the right side of the diagram (bordered in Fig.~\ref{fig01})
contributes to kinetics. In this case we consider only the equations
for conduction band electrons excluding the donor capture
\cite{Suppl}:

\begin{equation}
\frac{dn_\uparrow}{dt}=-\frac{n_\uparrow-n_\downarrow}{2\tau_s^e}-\gamma^{\rm
eA} n_\uparrow m^0_{\Uparrow\downarrow},
\end{equation}
\begin{equation}
\frac{dn_\downarrow}{dt}=\frac{\alpha^-}{M^-} m^-_\Downarrow
G-\frac{n_\downarrow-n_\uparrow}{2\tau_s^e}-\gamma^{\rm eA}
n_\downarrow m^0_{\Downarrow\uparrow}.
\end{equation}

It is easy to find an analytical result for the high pump power in
the limiting case of the fast ${\rm Mn^0_{Ga}}$ spin relaxation
($m^0_{\Uparrow\downarrow}=m^0_{\Downarrow\uparrow}=n/2$). For the
steady state the following result for the total concentration
$n=n_\uparrow+n_\downarrow$ and for a concentration difference
$\Delta n=n_\uparrow-n_\downarrow$ can be found:
\begin{equation}
n=M^-,\qquad \Delta n=-\frac{\gamma^{\rm
eA}(M^-)^2\tau_s^e}{\gamma^{\rm eA}M^-\tau_s^e+1},
\end{equation}
restoring the negative ($\Delta n<0$) magnetization that corresponds
to a spin alignment in accordance with the selection rules.

The full kinetic picture describing the transition from the positive
to the negative magnetization requires both excitation channels of
Fig.~\ref{fig01}. The magnetization sign depends on a relative
contribution of the different excitation channels as well as the
competition between the recombination channels (eA and DA
recombination). Thus, the concentration dependence of the absorption
coefficients and the bimolecular recombination rate lead to a
dependence of the electron magnetization on the excitation power. In
this case all 11 rate equations are required~\cite{Suppl}.

\begin{figure}
\includegraphics[width=86mm]{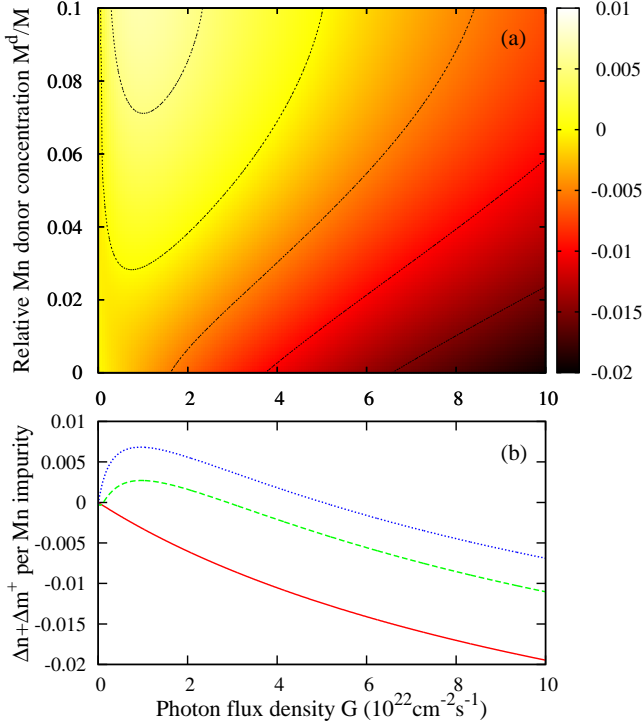}
\caption{\label{fig02} Dependence of the total electron
magnetization $(\Delta n+\Delta m^+)/M$ on the relative donor
concentration $M^d/M$ and the exciting photon flux density $G$. (a)
The total map in $G$ and $M^d$ axes. The magnetization isolines
[$(\Delta n+\Delta m^+)/M=-0.015, -0.01, -0.005, 0, +0.005$] are
depicted by thin dotted lines. (b) The cross sections of panel (a)
by lines $M^d={\rm const}$ at $M^d/M=0, 0.05, 0.1$ (solid red,
dashed green, and dotted blue line, respectively); $\tau_s^e=10$ ns,
$\tau_s^-=100$ ns, $\tau_s^0=1$ ns, $\tau_s^+=10$ ns,
$\tau_s^{2+}=200$ ns, $\gamma^{\rm eA}=1.2\times 10^{-10}$
cm$^3$s$^{-1}$, $\gamma^{\rm DA}=0.6\times 10^{-10}$ cm$^3$s$^{-1}$,
$\beta^{\rm sc}=10^{-8}$ cm$^3$s$^{-1}$, $\beta^{\rm sf}=2\times
10^{-9}$ cm$^3$s$^{-1}$, $\delta^{\rm sc}=10^{-9}$ cm$^3$s$^{-1}$,
$\delta^{\rm sf}=0$, $\alpha^d/M^d=3.0\times10^{-15}$ cm$^2$,
$\alpha^-/M^-=1.2\times 10^{-16}$ cm$^2$, $M^{\rm back}=0.7M$,
$M=10^{18}$ cm$^{-3}$.}
\end{figure}

The results of a numerical solution are depicted in
Fig.~\ref{fig02}, with the dependence of the total electron (both
free and localized) polarization $\Delta n+\Delta m^+$ being plotted
as a function of photon flux density and the relative Mn-donor
concentration. In the total kinetic picture we also use the
following spin-relaxation times, $\tau_s^0$ and $\tau_s^-$ for ${\rm
Mn_{Ga}^0}$ and ${\rm Mn_{Ga}^-}$, respectively. One can see the SRM
phenomenon as a function of the excitation power at a nonzero donor
concentration. The magnitude of the anomalous (positive)
magnetization increases with an increase of the donor concentration
$M^d$; the pump power corresponding to the change of the
magnetization sign increases as well. The relative electron
magnetization $(\Delta n+\Delta m^+)/(n+m^+)$ is higher than the
plotted value ($\Delta n+\Delta m^+)/M$, especially at the low pump
power when the electron concentration is low, $n+m^+\ll M$. The
magnitude of the effect is not discussed here; only the possibility
to invert the electron magnetization by means of the excitation
power variation is predicted.

The set of parameters for GaAs with the Mn concentration about
$10^{18}$cm$^{-3}$ is used for the numerical solution. The typical
value $\tau_s^e=10$ ns \cite{Astakhov2008} is utilized for the
conduction band electron spin-relaxation time. The absorption
coefficients are close to those of the impurity-to-band transition
\cite{Eagles1960}. The recombination parameters are of the order of
the magnitude known from the literature as well. The estimate of the
$\beta^{\rm sc}$ parameter as $2.7\times 10^{-8}$ cm$^3$s$^{-1}$ is
made based on a result of Ref.~\onlinecite{Bimberg1985}, where the
capture cross section $\sigma=5.1\times 10^{-15}$ cm$^2$ by a
shallow nonmagnetic donor was measured. The electron thermal
velocity $v_T=\sqrt{3k_BT/m^*}=5.2\times 10^6$cm/s for $T=4$ K and
GaAs electron effective mass $m^*=0.067m_0$ are used as well;
$\beta^{\rm sc}=\sigma v_T$. The capture parameter is assumed to be
independent on the magnetic nature of the donor. At the low
excitation intensity and temperature the captured electron flux is
usually higher than the eA recombination one. This is confirmed by
the well-known observation of the higher PL intensity of the DA band
compared to the eA band. Thus, the relation $\beta^{\rm
sc},\beta^{\rm sf}\gg\gamma^{\rm eA}$ is usually fulfilled. The
spin-relaxation time $\tau_s^{2+}$ is assumed to be the longest one,
since there is no bound charge carrier at ${\rm Mn^{2+}_I}$ center,
whereas, the relaxation of the internal $d$-shell spin provided by
the spin-lattice mechanism is slow. The parameters $\delta^{\rm
sc}$, $\delta^{\rm sf}$ are free parameters of our model. However,
these parameters as well as spin-relaxation times $\tau^-_s$,
$\tau_s^0$, and $\tau_s^+$ do not affect the SRM effect since they
do not enter Eqs.~(\ref{m_up}), and (\ref{m_down}) and the
conditions under which these equations are satisfied.

Since the time-dependent system of rate equations includes the
coupling between different charge states and Mn complexes, then the
magnetization switching time is determined by the longest
spin-relaxation time in the system. Thus, the switching time is
about $\tau_s^{2+}\sim 100$ ns. It should be noted that the temporal
behavior of the electron magnetization is nonmonotonic when the
system tends to the steady state.

\textit{Discussion of the experimental SRM observation.} In
practical terms, for the PL excitation spot area $S$ of
$10^{-9}-10^{-8}$ cm$^2$ and photon energy of 1.5 eV,
Fig.~\ref{fig02} corresponds to the absorbed pump power $P$ of a few
tens of microwatts. The experimentally observable quantity
characterizing the carrier spin polarization is a degree of the PL
circular polarization
\begin{equation}
\label{PL_polarization} \mathcal{P}=\frac{I^+-I^-}{I^++I^-},
\end{equation}
where $I^+$ ($I^-$) is the PL intensity with $\sigma^+$ ($\sigma^-$)
light polarization. The PL includes both eA- and DA-recombination
contribution and for PL-intensity with $\sigma^+$ ($\sigma^-$)
polarization one can write
\begin{equation}
\label{I_DA} I_{\rm DA}^+\propto\gamma^{\rm
DA}m^+_{\Downarrow\downarrow}m^0_{\Downarrow\uparrow},\qquad I_{\rm
DA}^-\propto\gamma^{\rm
DA}m^+_{\Uparrow\uparrow}m^0_{\Uparrow\downarrow},
\end{equation}
\begin{equation}
\label{I_eA} I_{\rm eA}^+\propto\gamma^{\rm eA}
n_{\downarrow}m^0_{\Downarrow\uparrow},\qquad I_{\rm
eA}^-\propto\gamma^{\rm eA}n_{\uparrow}m^0_{\Uparrow\downarrow},
\end{equation}
which are the selection rules for radiative recombination. The
position of the DA line in the PL spectra relative to the eA line
corresponds to the difference of the magnetic donor binding energy
(e.g., for the ${\rm Mn^+_I}$ donor in GaAs this energy does not
exceed a few meV) and the Coulomb attraction energy in the final
state. Thus, the two lines can be unresolved, for which reason we
consider the integral polarization of the PL band containing both eA
and DA lines.

\begin{figure}
\includegraphics[width=72mm]{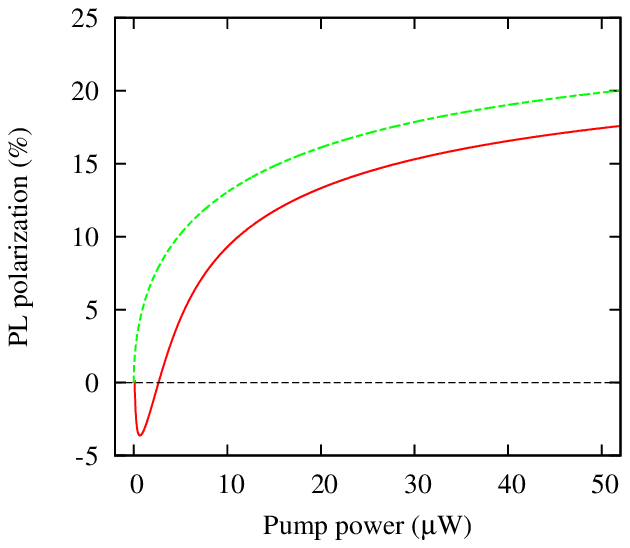}
\caption{\label{fig03} Dependence of the PL polarization involving
${\rm Mn^0}$-related optical transitions on the pump power.
Parameters are the following: $M=10^{18}$cm$^{-3}$, $S=10^{-9}$
cm$^2$, $\tau_s^e=10$ ns, $\tau_s^0=10$ ns,
$\tau_s^-=\tau_s^+=\tau_s^{2+}=100$ ns, $\gamma^{\rm DA}=6\times
10^{-11}$ cm$^3$s$^{-1}$, $\gamma^{\rm eA}=8\times 10^{-11}$
cm$^3$s$^{-1}$, $\delta^{\rm sc}=10^{-9}$ cm$^3$s$^{-1}$,
$\delta^{\rm sf}=0$, $\beta^{\rm sc}=3.0\times 10^{-8}$
cm$^3$s$^{-1}$, $\beta^{\rm sf}=1.5\times 10^{-8}$ cm$^3$s$^{-1}$,
$\alpha^-=100$ cm$^{-1}$, $\alpha^d=400$ cm$^{-1}$. Solid red curve
corresponds to the SRM behavior, $M^d=0.05M$, $M^{\rm back}=0.7M$;
dashed green curve corresponds to the normal behavior, $M^d=0$,
$M^{\rm back}=M$.}
\end{figure}

The calculated dependence of the integral PL polarization as a
function of the pump power is plotted in Fig.~\ref{fig03}. It has
been found by the numerical solution of the system of 11 rate
equations \cite{Suppl} and utilizing
Eqs.~(\ref{PL_polarization})--(\ref{I_eA}). Qualitatively the total
electron spin sign as well as the PL polarization degree depend on a
competition between the two above-mentioned channels of the
excitation and the recombination. The negative PL polarization at
the low pump power corresponds to the predominance of the DA
recombination. At the high excitation level the channel via donors
is saturated and the sign of the integral PL polarization becomes
positive as the eA recombination takes over. The phenomenon is more
pronounced for the higher DA-dimer concentration. Nevertheless, the
predicted SRM effect persists at any nonzero dimer concentration.
The $\mathcal{P}(P)$ dependence is monotonic as shown in
Fig.~\ref{fig03}, if the magnetic donor concentration is zero
($M^d=0$) only.

The estimates for the PL polarization in low/high pump power limits
are made. At the fast ${\rm Mn^0_{Ga}}$ spin relaxation
($\tau_s^0\rightarrow 0$, the PL polarization reflects electron
magnetization only) the degree of the circular PL polarization in
low pump limit ($\mathcal{P}=-\Delta m^+/m^+$) is negative:
\begin{equation}
\mathcal{P}_{\rm low}=-\tau_s^+\sqrt{\frac{\alpha^d\gamma^{\rm
DA}G}{2}}.
\end{equation}

In the high pump power limit the PL polarization degree is imposed
by the eA recombination, $\mathcal{P}=-\Delta n/n$. Due to the
predominant spin-down magnetization of the conduction band electrons
it is positive and saturated:
\begin{equation}
\label{P_high} \mathcal{P}_{\rm
high}=\frac{1}{1+\tau_l^\infty/\tau_s^e},
\end{equation}
with $\tau_l^\infty=1/(\gamma^{\rm eA}M^-)$ being the radiative
lifetime at the high pump power level. The high pump power behavior
resembles the well-known result for the PL polarization
\cite{Meier1984}. This formula is similar for the quantum-well case
under the excitation/recombination with the participation of the
first heavy-hole subband (in the case of the interband
excitation/recombination in the bulk ${\rm A^{III}B^V}$ material it
has to contain prefactor 0.25 as follows from the selection rules).
In the present case it is a consequence of a simplification imposed
by the pseudospin-1/2 model.

\textit{Conclusion.} As shown by the proposed consideration, the
magnetization in spin subsystems does not depend on the optical
selection rules only. The charge-carrier capture by the local
defects or impurities, the character of the spin-spin interactions,
and the spin relaxation are crucial for the optical orientation as
well. The all-optical scheme of the electron sign-reversal
magnetization in Mn-doped GaAs-based semiconductor structures has
been suggested and extensively studied. We also discuss an
experimental way to detect the SRM effect by means of the polarized
PL.

The switching of the electron magnetization by means of the
temperature control is also possible. Let us consider the above
system at the liquid-helium temperature and under the pump power
corresponding to the maximal positive magnetization. In this case
the magnetization is due to the donor alignment. The growing
temperature decreases the capture rate by the shallow donors and in
turn the anomalous magnetization is decreased as well. The energy
corresponding to the temperature of the liquid nitrogen is about 7
meV, while the electron binding energy of the ${\rm Mn_I^+}$ state
in GaAs is estimated as not exceeding 4 meV. The conduction band
electrons with such an energy are not captured by donors and
predominantly remain in the spin-down state. This means that at the
liquid-nitrogen temperature the negative magnetization is almost
completely restored.

\textit{Acknowledgments.} We are grateful to I.V. Rozhansky for
useful discussions. I.A.K. and N.S.A. acknowledge financial support
from the Russian Science Foundation (Grant No.~17-12-01182-c).

%----------------------------------------------


\newpage

%----------------------------------------------

\begin{widetext}

\section{Supplemental material: Minimal system of rate equations for the sign-reversal electron magnetization}

In the main text the kinetic processes within the pseudo-spin 1/2
model were described in words. All strength parameters and the
concentration designation were done as well. Here rate equations are
presented explicitly. For the conduction band electrons the
following pair of rate equations is written:

\begin{equation}
\frac{dn_\uparrow}{dt}=-\frac{n_\uparrow-n_\downarrow}{2\tau_s^e}-\gamma^{\rm
eA}n_\uparrow m^0_{\Uparrow\downarrow}-\beta^{\rm sc}n_\uparrow
m^{2+}_\Uparrow-\beta^{\rm sf} n_\uparrow m^{2+}_\Downarrow.
\end{equation}

Here the terms in the right hand side describe the electron spin
relaxation, eA-recombination, and the electron capture by donors
with the spin conservation or the spin flip, respectively. The rate
equation for the electron concentration at the opposite spin state
is derived by reversing spin indices. Additionally, here two terms,
corresponding to the $\sigma^+$-excitation from the ionized acceptor
state or from the DA-dimer one, arise in accordance with selection
rules:

\begin{equation}
\frac{dn_\downarrow}{dt}=\frac{\alpha^-}{M^-}m^-_\Downarrow G+
\frac{\alpha^d}{M^d}m^dG-\frac{n_\downarrow-n_\uparrow}{2\tau_s^e}-\gamma^{\rm
eA}n_\downarrow m^0_{\Downarrow\uparrow}-\beta^{\rm sc}n_\downarrow
m^{2+}_\Downarrow-\beta^{\rm sf}n_\downarrow m^{2+}_\Uparrow.
\end{equation}

The neutral ${\rm Mn_{Ga}^0}$ acceptors can be at equilibrium inside
the system (depending on the compensation degree) and can arise at
the photoexcitation. The processes similar to those of the electrons
can be observed here. However, an additional term occurs that
describes the DA-recombination,

\begin{equation}
\frac{dm^0_{\Uparrow\downarrow}}{dt}=-\frac{m^0_{\Uparrow\downarrow}-m^0_{\Downarrow\uparrow}}{2\tau_s^0}-\gamma^{\rm
eA}m^0_{\Uparrow\downarrow}n_\uparrow- \gamma^{\rm
DA}m^0_{\Uparrow\downarrow}m^+_{\Uparrow\uparrow},
\end{equation}
\begin{equation}
\frac{dm^0_{\Downarrow\uparrow}}{dt}=\frac{\alpha^-}{M^-}m^-_\Downarrow
G+
\frac{\alpha^d}{M^d}m^dG-\frac{m^0_{\Downarrow\uparrow}-m^0_{\Uparrow\downarrow}}{2\tau_s^0}-\gamma^{\rm
eA}m^0_{\Downarrow\uparrow}n_\downarrow -\gamma^{\rm
DA}m^0_{\Downarrow\uparrow}m^+_{\Downarrow\downarrow}.
\end{equation}

A pair of the rate equations for the singly charged Mn donors ${\rm
Mn^+_I}$ is similar to above one:
\begin{equation}
\frac{dm^+_{\Uparrow\uparrow}}{dt}=-\frac{m^+_{\Uparrow\uparrow}-m^+_{\Downarrow\downarrow}}{2\tau_s^+}-\gamma^{\rm
DA}m^+_{\Uparrow\uparrow}m^0_{\Uparrow\downarrow}+\beta^{\rm
sc}m^{2+}_\Uparrow n_\uparrow+\beta^{\rm sf}m^{2+}_\Uparrow
n_\downarrow,
\end{equation}
\begin{equation}
\frac{dm^+_{\Downarrow\downarrow}}{dt}=-\frac{m^+_{\Downarrow\downarrow}-m^+_{\Uparrow\uparrow}}{2\tau_s^+}-\gamma^{\rm
DA}m^+_{\Downarrow\downarrow}m^0_{\Downarrow\uparrow}+\beta^{\rm
sc}m^{2+}_\Downarrow n_\downarrow+\beta^{\rm sf}m^{2+}_\Downarrow
n_\uparrow.
\end{equation}

The doubly charged donors ${\rm Mn_I^{2+}}$ arise in our scheme as a
result of either the DA-dimer decay or the DA-recombination. The
spin relaxation and the concentration decrease due to the donor
capture are taken into account as well. The processes of the dimer
formation are considered in the bimolecular way. We suppose that the
DA-dimer formation without spin-flip is more effective, $\delta^{\rm
sc}\gg\delta^{\rm sf}$. Moreover, at numerical calculations we use
$\delta^{\rm sf}=0$ because we take into account the finite spin
relaxation times both for ${\rm Mn^{2+}_I}$ and ${\rm Mn^-_{Ga}}$
(see below), that gives the possibility of the ground state dimer
formation with the conservation of d-shell spins. The rate equation
for ${\rm Mn^{2+}_I}$ are given by

\begin{equation}
\frac{dm^{2+}_\Uparrow}{dt}=
\frac{\alpha^d}{M^d}m^dG-\frac{m^{2+}_\Uparrow-m^{2+}_\Downarrow}{2\tau_s^{2+}}+\gamma^{\rm
DA}m^+_{\Uparrow\uparrow}m^0_{\Uparrow\downarrow}-\beta^{\rm
sc}m^{2+}_\Uparrow n_\uparrow-\beta^{\rm sf}m^{2+}_\Uparrow
n_\downarrow-\delta^{\rm sc}m^{2+}_\Uparrow
m^-_\Downarrow-\delta^{\rm sf}m^{2+}_\Uparrow m^-_\Uparrow,
\end{equation}
\begin{equation}
\frac{dm^{2+}_\Downarrow}{dt}=-\frac{m^{2+}_\Downarrow-m^{2+}_\Uparrow}{2\tau_s^{2+}}+\gamma^{\rm
DA}m^+_{\Downarrow\downarrow}m^0_{\Downarrow\uparrow}-\beta^{\rm
sc}m^{2+}_\Downarrow n_\downarrow-\beta^{\rm sf}m^{2+}_\Downarrow
n_\uparrow-\delta^{\rm sc}m^{2+}_\Downarrow m^-_\Uparrow-\delta^{\rm
sf}m^{2+}_\Downarrow m^-_\Downarrow.
\end{equation}

In the right hand side of equations describing the ionized ${\rm
Mn_{Ga}^-}$-acceptor population the above-mentioned processes
(excitation, spin relaxation, eA- and DA-recombination and dimer
formation) are included:
\begin{equation}
\frac{dm^-_\Uparrow}{dt}=-\frac{m^-_\Uparrow-m^-_\Downarrow}{2\tau_s^-}+\gamma^{\rm
eA}n_\uparrow m^0_{\Uparrow\downarrow}+\gamma^{\rm
DA}m^+_{\Uparrow\uparrow}m^0_{\Uparrow\downarrow}-\delta^{\rm
sc}m^-_\Uparrow m^{2+}_\Downarrow-\delta^{\rm sf}m^-_\Uparrow
m^{2+}_\Uparrow,
\end{equation}
\begin{equation}
\frac{dm^-_\Downarrow}{dt}=-\frac{\alpha^-}{M^-}m^-_\Downarrow
G-\frac{m^-_\Downarrow-m^-_\Uparrow}{2\tau_s^-}+\gamma^{\rm
eA}n_\downarrow m^0_{\Downarrow\uparrow}+\gamma^{\rm
DA}m^+_{\Downarrow\downarrow}m^0_{\Downarrow\uparrow} -\delta^{\rm
sc}m^-_\Downarrow m^{2+}_\Uparrow -\delta^{\rm sf}m^-_\Downarrow
m^{2+}_\Downarrow.
\end{equation}

The change of the DA-dimer concentration $m^d$ is a result of the
competition between the photo-excitation process accompanied by a
decay of the dimer and the above-mentioned process of the dimer
formation:

\begin{equation}
\frac{dm^d}{dt}=-\frac{\alpha^d}{M^d}m^dG+\delta^{\rm
sc}(m^-_\Uparrow m^{2+}_\Downarrow+m^-_\Downarrow
m^{2+}_\Uparrow)+\delta^{\rm sf}(m^-_\Uparrow
m^{2+}_\Uparrow+m^-_\Downarrow m^{2+}_\Downarrow).
\end{equation}

It should be noted, that these 11 equations are not linearly
independent. The additional restrictions have to be imposed in order
to couple variables. The additional equations can be found by making
the linear combination of some rate equations giving the zero right
hand side, $\sum a_i\frac{d...}{dt}=0$, with $a_i$ being the
dimensionless constants. This means that a linear combination of
variables (concentrations) is constant (time-independent). The
magnitude of this constant can be found from the initial conditions
at $G=0$.

In this manner three additional equations in the form of the
`conservation laws' can be found. The first one corresponds to the
conservation of the total Mn concentration in all possible charge
states:
\begin{equation}
\label{conserve1}
m^0_{\Uparrow\downarrow}+m^0_{\Downarrow\uparrow}+m^+_{\Uparrow\uparrow}+m^+_{\Downarrow\downarrow}+m^{2+}_\Uparrow+m^{2+}_\Downarrow+m^-_\Uparrow+m^-_\Downarrow+2m^d=M.
\end{equation}

The second one describes the charge conservation in the system,
i.e., the total concentration of the photoexcited electrons (both
free and localized ones) has to be equal to the concentration of the
photoexcited ${\rm Mn^0_{Ga}}$ holes:
\begin{equation}
\label{conserve2}
n_\uparrow+n_\downarrow+m^+_{\Uparrow\uparrow}+m^+_{\Downarrow\downarrow}=m^0_{\Uparrow\downarrow}+m^0_{\Downarrow\uparrow}-M^0.
\end{equation}
Before illumination we have $M^0=M-M^{\rm back}-3M^d$. The factor 3
reflects the facts that ${\rm Mn_I}$ is a double donor, and in
addition, after the ionization of two Mn acceptors, each Mn donor is
involved in the formation of the Mn-dimer.

The third 'conservation law' corresponds to the conservation of the
total Mn-donor concentration:
\begin{equation}
\label{conserve3}
m^+_{\Uparrow\uparrow}+m^+_{\Downarrow\downarrow}+m^{2+}_\Uparrow+m^{2+}_\Downarrow+m^d=M^d.
\end{equation}

In this manner the 'conservation law' can be written for the total
Mn-acceptor concentration. However, it is not linear-independent
one, as there are such equations for the Mn donors and for the total
Mn concentration (donors+acceptors). This equation can be simply
derived by subtracting Eq.~(\ref{conserve3}) from
Eq.~(\ref{conserve1}).

\end{widetext}

\end{document}